\documentclass[aps,prl,floatfix,nofootinbib,letterpaper,twocolumn,superscriptaddress]{revtex4}

\usepackage{graphics,graphicx,epsfig,bbm}
\usepackage{amsmath,amsfonts,amssymb,amscd,latexsym,color}
\usepackage{amsthm}

\usepackage{times}
\usepackage{pstricks}

\newcommand{\1}{\openone}

\newcommand{\comp}{\circ}

\newcommand{\R}{{\mathbb R}}

\newcommand{\tr}{{\mathrm{tr}}}

\newcommand{\tLambda}{{\bar{\Lambda}}}
\newcommand{\ptLambda}{{\bar{\Lambda}_\Pi}}

\newcommand{\w}{\mathbf{w}}

\newcommand{\kket}[1]{|{#1}\rangle\!\rangle}
\newcommand{\bbra}[1]{\langle\!\langle{#1}|}

\DeclareMathAlphabet{\matheu}{U}{eus}{m}{n}
\newcommand{\Paulis}{{\matheu{P}_n}}

\begin{document}

\title{Experimentally scalable protocol for identification of
  correctable codes}

\author{M. Silva}
\affiliation{
Institute for Quantum Computing,
University of Waterloo,
Waterloo, Ontario, Canada, N2L 3G1}
\affiliation{
Department of Physics and Astronomy,
University of Waterloo,
Waterloo, Ontario, Canada, N2L 3G1}
\author{E. Magesan}
\affiliation{
Institute for Quantum Computing,
University of Waterloo,
Waterloo, Ontario, Canada, N2L 3G1}
\affiliation{
Department of Applied Mathematics,
University of Waterloo,
Waterloo, Ontario, Canada, N2L 3G1}
\author{D. W. Kribs}
\affiliation{
Institute for Quantum Computing,
University of Waterloo,
Waterloo, Ontario, Canada, N2L 3G1}
\affiliation{
 Department of Mathematics and Statistics,
University of Guelph,
Guelph, Ontario, Canada, N1G 2W1 }
\author{J. Emerson}
\affiliation{
Institute for Quantum Computing,
University of Waterloo,
Waterloo, Ontario, Canada, N2L 3G1}
\affiliation{
Department of Applied Mathematics,
University of Waterloo,
Waterloo, Ontario, Canada, N2L 3G1}

\date{\today}

\begin{abstract}

  The task of finding a correctable encoding that protects against
  some physical quantum process is in general hard. Two main obstacles
  are that an exponential number of experiments are needed to gain
  complete information about the quantum process, and known
  algorithmic methods for finding correctable encodings involve
  operations on exponentially large matrices. However, we show that in
  some cases it is possible to find such encodings with only partial
  information about the quantum process.  Such useful partial
  information can be systematically extracted by averaging the channel
  under the action of a set of unitaries in a process known as {\em
    twirling}.  In this paper we prove that correctable encodings for
  a twirled channel are also correctable for the original channel. We
  investigate the particular case of twirling over the set of Pauli
  operators and qubit permutations, and show that the resulting
  quantum operation can be characterised experimentally in a scalable
  manner. We also provide a postprocessing scheme for finding
  unitarily correctable codes for these twirled channels which does
  not involve exponentially large matrices.

\end{abstract}

\maketitle


{\bf\em Introduction ---} The coherent experimental manipulation of
quantum systems, and their application to various quantum information
tasks, confronts significant limitations in the presence of noise, and
in particular, decoherence.  The discovery of quantum error correction
codes enables methods for overcoming these limitations whenever the
decoherence satisfies various well-defined sets of
conditions. Specifically, the appliciability of particular codes
depends crucially on the detais of the physical noise model affecting
a particular system. The standard approach for experimentally
characterizing the full noise model and then assessing the usefulness
of a given code are costly procedures, in particular, requiring a
number of experiments~\cite{chuang97,mohseni07} that grows
exponentially with the number of subsystems and the (classical)
post-processing of matrices~\cite{HKL04,codesearch} whose dimension
grows similarly exponentially. These limitations make the standard
approach infeasible for the kinds of quantum information systems
required in practice. In this paper we describe a method that
overcomes both of these issues.

First, we demonstrate that {\em partial information about the noise is
sufficient to construct correctable encodings}.  This is motivated
by the recent demonstration that valuable partial information about
the noise can be extracted by symmetrising or ``twirling'' the quantum
channel~\cite{twirl,dankert06,emerson07}. Twirling consists of averaging the
action of a sequence of gates over some distribution of unitaries, in
effect randomising aspects of the channel. This leads to
scalable protocols which answer questions of practical interest about
the noise. In particular, it is possible to experimentally estimate
the average gate fidelity between a quantum operation and the identity
map~\cite{dankert06}, as well as to obtain information about noise
correlations in a quantum process~\cite{emerson07}.

We then focus on a particular type of twirl, the {\em Pauli twirl},
which has a number of special features that make finding noiseless and
unitarily correctable encodings more tractable than the general
case. First, the partial information obtained via a Pauli twirl can be
accessed with only a polynomial number of experiments, making such
experimental characterisation strictly scalable. Then, we discuss how
this partial information lends itself to an algebraic algorithm for
finding correctable encodings. This new approach does not require the
manipulation of exponentially large matrices. We also discuss how
these encodings can be verified experimentally in an efficient manner.


{\bf\em Twirled Quantum Operations ---} Consider a trace preserving
linear quantum operation $\Lambda(\rho)=\sum_{i,j} [\chi]_{ij} P_i\rho
P_j$ acting on $n$ qubits, where the $P_{i,j}\in\Paulis$ are all Pauli
operators acting acting on $n$ qubits.  Trace-preservation requires
that $\sum_{i,j} [\chi]_{ij} P_j P_i = \1$ (in particular, $\tr
\chi=1$), while Hermiticity-preservation requires
$[\chi]_{ij}=[\chi]_{ji}^*$. Completely-positive (CP) operations have
the aditional requirement that $\chi \ge 0$, which follows from the
fact that CP maps can be written as a Choi-Kraus sum
$\Lambda(\rho)=\sum_{k}A_k\rho A_k^\dagger$~\cite{choikraus}.  This
implies that for CP maps, $\forall i \quad [\chi]_{ii}\ge 0$, so the
$[\chi]_{ii}$ can be interpreted as probabilities.  

Quantum operations require $O(2^{4n})$ parameters to be fully
described, and thus require an exponential number of experiments to be
fully characterised~\cite{chuang97,mohseni07}.  Due to this
exponential cost, it is impractical to obtain a complete description
about noise and decoherence acting on even a moderately large system
of qubits.  Useful partial information about the noise can be obtained
by averaging the action of the quantum operation under the composition
${\mathcal U} \comp \Lambda \comp {\mathcal U}^\dagger$ for unitary
operations ${\mathcal U}(\rho)=U\rho U^\dagger$ randomly chosen
according to some distribution~\cite{dankert06,emerson07}. This
averaging is known as a ``twirl'', and the averaged channel
$\tLambda(\rho) = \int d\mu({\mathcal U}) {\mathcal U} \comp \Lambda
\comp {\mathcal U}^\dagger(\rho)$ is known as the ``twirled
channel''. The case where the distribution over unitaries is discrete
is of particular interest. In that case, the twirled channel is given by
$\tLambda(\rho) = \sum \Pr({\mathcal U}_i) {\mathcal U}_i \comp \Lambda \comp
{\mathcal U}_i^\dagger(\rho)$, where $\Pr({\mathcal U}_i)$ is a probability
distribution over the ${\mathcal U}_i$. This leads us to our first result.

{\em Theorem :} Any correctable code for a twirled channel
$\tLambda$ is a correctable code for the original channel $\Lambda$ up
to an aditional unitary correction.

Proof: Without loss of generality, consider
$\Lambda(\rho)=\sum_{k}A_k\rho A_k^\dagger$ and a twirl with unitaries
$\{U_j\}$ where $U_1=\1$. Any unitary twirl is unitarily equivalent to
a twirl that includes the identity, and this unitary equivalence leads
to the aditional unitary correction. A set of Choi-Kraus operators for
$\tLambda$ is then $\{U_j^\dagger A_kU_j\}$. The existence of a
correctable code for ${\mathcal H}_A$ under the action of $\tLambda$,
with recovery operations ${\mathcal R}(\rho)=\sum_m R_m\rho
R_m^\dagger$, corresponds to a projector $P$ into a subspace of the
form ${\mathcal H}_A\otimes{\mathcal H}_B$ such that $PR_mU_j^\dagger
A_kU_jP=R_mU_j^\dagger A_kU_jP$ and $R_mU_j^\dagger
A_kU_j|_P\in\1_A\otimes\mathcal{B}({\mathcal H}_B)$ is satisfied for
all $j,k, m$~\cite{kribs05}. Since $U_1=\1$, this implies $PR_m
A_kP=R_m A_kP$ and $R_m A_k|_P \in\1_A\otimes\mathcal{B}({\mathcal
  H}_B)$ for all $k,m$. Thus ${\mathcal H}_A$ is also correctable
under the action of $\Lambda$.\qed

We now consider some specific twirls which are known to yield
twirled channels which can be characterized efficiently via
experiments.


{\bf\em Pauli Twirling ---} 
It has been demonstrated~\cite{dankert06} that twirling
a channel $\Lambda$ by the Pauli group $\Paulis$ yields
the effective channel $\tLambda$ of the form
\begin{equation}
\tLambda(\rho)={1\over 4^n}\sum_{P_i\in\Paulis} P_i \Lambda( P_i \rho P_i ) P_i
= \sum_{i} [\chi]_{ii} P_i\rho P_i.
\end{equation}
In other words, the off-diagonal elements of $\chi$ are
eliminated. Channels of this form are known as 
{\em Pauli channels}. From here on $\tLambda$ will denote the
result of Pauli twirling $\Lambda$, unless stated otherwise.

Pauli channels have a number of useful properties which facilitate the
search for some types of correctable codes. However, the description
of a general Pauli channel still requires an exponential number of
parameters, and such parameters are not realistically accessible due
to this exponential overhead. Instead, one can consider an additional
twirl by the group $\Pi_n$ consisting of all qubit permutations. The
channel $\ptLambda$ resulting from a combination of a $\Paulis$ twirl
and a $\Pi_n$ twirl is what we call a {\em permutation invariant
  Pauli} (PIP) channel. Such channels can be written in terms of
Choi-Kraus operators
\begin{equation}\label{kraus}
\ptLambda(\rho) = 
\sum_\w  p_\w 
    \sum_{\nu_\w} {1\over K^\nu_\w}\\
    \sum_{\mathbf{i}_\w}
      {1\over K^{\mathbf{i}}_{\nu_\w}}
    P_{w,\nu_\w,\mathbf{i}_\w} \rho P_{w,\nu_\w,\mathbf{i}_\w},
\end{equation}
where $\mathbf{w}=(w_x,w_y,w_z)$ labels the number of the
X, Y and Z Pauli operators, 
$\nu_\w$ labels the ${w_x+w_y+w_z}$ qubits over which 
$P_{\w,\nu_\w,\mathbf{i}_\w}$ acts non-trivially, 
and $\mathbf{i}_\w$
labels which single qubit Pauli operator act on each of 
the qubits. The number $K_n$ of different labels $\w$ needed to 
describe such a channel is
$ K_n\equiv {1 \over 6} n^3 + n^2 + {11\over 6} n + 1,$
while for a fixed $\w$ there are
$K^\nu_\w={n \choose w_x+w_y+w_z}$ different $\nu_\w$,
and for fixed $\w$ and $\nu_\w$, there are
$K^{\mathbf{i}}_{\nu_\w}={w_x+w_y+w_z\choose w_x}{w_y+w_z\choose w_z}$ 
different $\mathbf{i}_\w$.

Two equivalent descriptions of the channel (with the same
number of parameters) are the Choi-Kraus decomposition
and the diagonal representation. 
The Choi-Kraus decomposition \eqref{kraus} can be rewritten as
$\ptLambda(\rho) = \sum_{\w} p_\w M^p_\w(\rho),$ 
where $M^p_\w$ are the superoperators
\begin{equation}
M^p_\w(\rho) = 
{1\over K^\nu_\w} 
\sum_{\nu_\w}  {1\over K^{\mathbf{i}}_{\nu_\w}} \sum_{\mathbf{i}_\w}
P_{\w,\nu_\w,{\bf i}_\w} \rho P_{\w,\nu_\w,{\bf i}_\w}.
\end{equation}
The $M^p_\w$ are trace-preserving channels 
which apply each of the $P_{\w,\nu_\w,{\bf i}_\w}$ for a given $\w$ with the
same probability. The $M^p_\w$ form a basis for PIP channels.

Note that, given some $n$-fold tensor product of Pauli operators
$P_{\w,\nu_\w,{\bf i}_\w}$, we have $\ptLambda(P_{\w,\nu_\w,{\bf
    i}_\w}) = \lambda_\w P_{\w,\nu_\w,{\bf i}_\w}$ for some real
constant $\lambda_\w\in[-1,1]$, since Pauli operators either commute
or anti-commute. Moreover $\ptLambda$ is self-dual,
i.e. $\ptLambda=\ptLambda^\dagger$, a fact that follows directly from
the definition of Pauli channels.  Since the $\{P_{\w,\nu_\w,{\bf
    i}_\w}\}$ form an operator basis, it follows that the $\lambda_\w$
are the eigenvalues of $\ptLambda$, with high degeneracy, as they
depend only on $\w$. Thus, Pauli channels are diagonalisable and
Hermitian.

If we define $\kket{A}\bbra{B} \rho \equiv {1\over 2^n} A~\tr(B^\dagger\rho)$
then the diagonal representation of the channel $\ptLambda$ 
in terms of its eigenoperators is $\ptLambda(\rho) = \sum_{\w=0}^n \lambda_\w M^\lambda_\w(\rho),$
where $M^\lambda_\w$ are the superoperators
\begin{equation}
M^\lambda_\w(\rho) 
 =
\sum_{\nu_\w}\sum_{\mathbf{i}_\w}
\kket{P_{\w,\nu_\w,{\bf i}_\w}}\bbra{P_{\w,\nu_\w,{\bf i}_\w}} \rho.
\end{equation}

The diagonal description of $\ptLambda$ 
allows for straighforward description of the composition of
channels. For example, suppose two Pauli channels $\ptLambda^{(1)}$ and $\ptLambda^{(2)}$
are be composed to yield $\ptLambda^{(1)}\comp\ptLambda^{(2)}$.  This simply
translates to the multiplication of the corresponding eigenvalues for each
of the two channels because all Pauli channels commute. 

It is crucial to note that, because there is at most a polynomial
number of different eigenvalues, they can be estimated efficiently by
determining how a Pauli observable in a $\w$ class is scaled under the
action of the twirled channel, as described in
Ref.~\cite{emerson07}. Thus, all the eigenvalues of a PIP channel can
be estimated experimentally in a scalable manner. 



{\bf\em Unitarily Correctable Subsystems ---} Let $A$ and $B$ be
subsystems of a Hilbert space $\mathcal H = {\mathcal H}_A \otimes
{\mathcal H}_B \oplus {\mathcal H}_{K}$. Given some channel $\Lambda :
\mathcal B (\mathcal H) \to \mathcal B (\mathcal H)$, we say that
${\mathcal H}_A$ is a {\em unitarily correctable
  subsystem}~\cite{KS06} if there is a unitary recovery operation
$\mathcal R$ acting on $\mathcal B(\mathcal H)$ such that
$\forall\sigma_A\in\mathcal B({\mathcal H}_A)$ and
$\forall\sigma_B\in\mathcal B({\mathcal H}_B)$ there exists a
$\tau_B\in\mathcal B({\mathcal H}_B)$ such that $\mathcal R \comp
\Lambda (\sigma_A \otimes \sigma_B) = \sigma_A \otimes \tau_B$. That
is, ${\mathcal H}_A$ defines a correctable encoding for $\Lambda$ that
can be returned to its initial location within the system Hilbert
space with a single unitary operation. The problem of finding
unitarily correctable subsystems for a unital channel is equivalent to
finding the structure of the commutant of the noise algebra of
$\Lambda^\dagger\comp\Lambda$~\cite{KS06}. In terms of the Choi-Kraus
operators $\{A_k\}$ for $\Lambda$, this ``noise commutant'' is defined
as the set of operators that commute with the operators $\{A_k^\dagger
A_j\}$. If $\Lambda$ is unital and trace preserving, this commutant
coincides with the {\em fixed point set} of
$\Lambda^\dagger\comp\Lambda$ (the set of operators which are
invariant under the action of this composed
channel)~\cite{HKL04}. This result can be refined somewhat for
diagonalisable channels.

{\em Proposition 1 ---} Let $\Lambda$ be a unital, diagonalisable and trace
preserving channel with eigenvalues $\lambda_i$ and eigenoperators
$L_i$.  Then, the noise commutant of $\Lambda^\dagger\comp\Lambda$ is the
space spanned by eigenoperators $L_i$ with eigenvalues
$|\lambda_i|=1$.

Proof: $\Lambda$ is unital and trace preserving, thus so is $\Lambda^\dagger\comp\Lambda$. 
Moreover, $\Lambda$ is diagonalisable, so
$\Lambda^\dagger\comp\Lambda$ has eigenoperators $L_i$ with
eigenvalues $\lambda_i^*\lambda_i=|\lambda_i|^2$, where the $\lambda_i$ are the
eigenvalues of $\Lambda$. Since $\Lambda^\dagger\comp\Lambda$ is unital, its
fixed point set and its noise commutant coincide, and both are
given by the space spanned by the eigenoperators $L_i$ with
$|\lambda_i|^2=1$.\qed

Pauli channels are unital channels, and since they  are
diagonalisable and Hermitian, these  channels have a particularly
simple fixed-point set structure. In particular, we immediatelly
obtain  the following result.

{\em Corollary 1 ---} Let $\tLambda$ be a Pauli channel. Then the noise
commutant of $\tLambda^\dagger\comp\tLambda$ is the space spanned by
the eigenoperators with eigenvalues $\pm1$.

Before continuing, we discuss a special class of UCS codes considered
in~\cite{KNPV07}. We say that a UCS code is {\em unitarily noiseless}
(UNS) for $\Lambda$ if it is a UCS of $\Lambda^n$ for all $n\geq 1$,
where $\Lambda^n$ is the channel $\Lambda$ composed with itself $n$
times. They are codes for which the recovery
operation has the special form $\mathcal U = \mathcal U_A \otimes
\mathcal R_B$, that is, a unitary acting only on the subsystem in
which information is preserved, and an arbitrary quantum channel on
subsystem $B$. Interestingly, the sets of UCS and UNS codes coincide
for Pauli channels.

{\em Corollary 2 ---} A UCS code of a Pauli channel $\tLambda$ is also a UNS.

Proof: As $\tLambda$ is Hermitian, it commutes with its dual
$\tLambda^\dagger=\tLambda$, and thus we have that
$\left(\tLambda^n\right)^\dagger\comp\tLambda^n=(\tLambda^\dagger\comp\tLambda)^n$.
Since the eigenvalues of $\tLambda$ are real, this implies that the
fixed point set of $\tLambda^\dagger\comp\tLambda$ is identical to the
fixed point set of
$\left(\tLambda^n\right)^\dagger\comp\tLambda^n$.\qed


{\bf\em PIP channel parameter space ---} 
Considering the Liouville representation of the 
superoperators $\{M^p_\w\}$ and $\{M^\lambda_\w\}$, 
it is easy to show that there is a linear invertible map 
$\Omega : \R^{K_n} \to \R^{K_n}$ mapping the ${p_\w}$ to the ${\lambda_\w}$.
More explicitly, we have that
\begin{equation}
\lambda_\w = \sum_{\mathbf v}[\Omega]_{\w,\mathbf v} p_{\mathbf v}\qquad
[\Omega]_{\w,\mathbf v} = 
{\langle M^\lambda_\w, M^p_{\mathbf v}\rangle \over \langle M^\lambda_\w, M^\lambda_\w\rangle},
\end{equation}
where $\langle \cdot, \cdot \rangle$ is the Hilbert-Schmidt
inner product in the Liouville representation. This follows
directly from the fact that 
$\langle M^\lambda_\w, M^\lambda_{\mathbf v}\rangle = \delta_{\w,{\mathbf v}}\langle M^\lambda_\w, M^\lambda_\w\rangle$. Similarly, because $\langle M^p_\w, M^p_{\mathbf v}\rangle = \delta_{\w,{\mathbf v}}\langle M^p_\w, M^p_\w\rangle$, we have
\begin{equation}
p_\w = \sum_{\mathbf v}[\Omega^{-1}]_{\w,\mathbf v} \lambda_{\mathbf v}\qquad
[\Omega^{-1}]_{\w,\mathbf v} = 
{\langle M^p_\w, M^\lambda_{\mathbf v}\rangle \over \langle M^p_\w, M^p_\w\rangle},
\end{equation}

The $\{p_\w\}$ form a $K_n-1$ simplex 
(the {\em probability 
simplex}, which is in fact the standard $K_n-1$ simplex), and
the $\{\lambda_\w\}$ also form a $K_n-1$ simplex, which we call
the {\em eigenvalue simplex}.

Explicitly computing the entries of the $\Omega$ matrix from the
matrix elements of Liouville representation of the superoperators is
inefficient, as these representations are exponentially large in
$n$. However, one can show that
$[\Omega]_{\mathbf{v},\w}=1-2{N(\mathbf{v},\w)\over4^n},$ where
$N(\mathbf{v},\w)$ is the number of Pauli operators in the Kraus
decomposition of an extremal channel (i.e. $p_{\mathbf v}=1$ for some
$\mathbf v$) which anticommute with some Pauli operator corresponding
to the equivalence class $\w$~\cite{emerson07}.  General expressions
for $N(\mathbf{v},\w)$ can be obtained by simple counting arguments,
which can be computed efficiently.

The fixed points of $\ptLambda$ can be thought of as the observables
that are conserved under the action of $\ptLambda$ in the Heisenberg
picture, as $\ptLambda$ is self-dual.  Once the fixed points of these
channels are determined, in order to determine possible encodings one
needs to compute the possible algebras generated by these fixed
points.  Note that, because of the degeneracy of the eigenvalues of a
PIP channel, the existence of a single weight class with eigenvalue
$1$ corresponds to a large number of Pauli operators which are fixed
points of the channel.  We now describe how this allows us to find
correctable codes for a $\ptLambda$.


{\bf\em Finding Correctable Codes ---} The identification of the fixed
point set of a unital channel can be used to find a UCS using the
general algorithm described in Ref.~\cite{HKL04,KS06}. However, this
algorithm requires the manipulation of exponentially large matrices, a
problem shared with numerical algorithms used to search for more
general codes~\cite{codesearch}.  Given the Pauli operators form an
eigenbasis for PIP channels, the task of computing the algebra of
conserved observables is relatively simpler than in the general case.
One simply has to group the conserved Pauli observables into triplets
of observables which satisfy the commutation relations for $su(2)$.
These commutation relations can be computed without writing the
observables explicitly in a particular representation. By choosing the
largest set $S$ of mutually exclusive triplets which commute with each
other, one implicitly describes how to encode a noiseless Hilbert
space of dimension $2^{|S|}$. Before discussing this algorithm, we
point out the following result, which can be obtained by direct
computation.

{\em Proposition 2} --- Given Pauli operators $\{P_j,P_k,P_l\}$
satisfying the commutation relations $[P_j,P_k]=2i\sum_l
\tilde{\epsilon}_{jkl}P_l$, where
$\tilde{\epsilon}_{jkl}=\epsilon_{jkl}{s_l\over s_js_k}$,
$\epsilon_{jkl}$ is the Levi-Civita symbol and $s_j,s_k,s_l\in\{\pm
1,\pm i\}$, then $\{s_jP_j,s_kP_k,s_lP_l\}$ obey the $su(2)$
commutation relations.

It is clear that $\{s_jP_j,s_kP_k,s_lP_l\}$ are unitarily related to
$\{X,Y,Z\}$, as all Pauli operators have the same spectrum and Proposition 2
guarantees they have the same commutation relations. The unitary which
performs the encoding is guaranteed to be in the Clifford group, since
it maps a set of Pauli operators to another set of Pauli operators
with the same commutation relations.  Standard techniques can be
applied to determine which Clifford group operations implement a
desired encoding~\cite{gottesman}.

This leads to the following algorithm for finding a correctable
encoding given the eigenvalues of the PIP channel $\tLambda$: (i)
Enumerate the Pauli operators with eigenvalues 1 under the action of
$\tLambda$, and call thise set $F$.  (ii) Choose a triplet of Pauli
operators satisfying the commutation relations in Proposition 2 -- if
none can be found, the search is over.  (iii) Remove this triplet from
$F$, as well as all operators that do not commute with the triplet,
and go back to step ii. The number of mutually exclusive triplets
found in this manner corresponds to an allowable number of encoded
qubits which can be protected from the action of $\ptLambda$.

Finding unitarily correctable subsystems is similarly simple. One can
easily compute the conserved observables of the channel
$\ptLambda^\dagger\circ\ptLambda$~\cite{KS06}.  In the case of PIP
channels, this corresponds to finding observables with eigenvalue $\pm
1$ for the channels $\ptLambda$, so that these observables have
eigenvalue $1$ for $\ptLambda^\dagger\circ\ptLambda=\ptLambda^2$. The
same procedure described above can be applied to the $\pm1$ eigenspace
to find a correctable encoding.


{\em Example 1} --- Consider the 2 qubit PIP channel with Kraus
operators proportial to $\{\1\1,ZZ\}$.  The Pauli operators with
eigenvalue $1$ are $\1\1, XX, YY, ZZ, XY, YX$, and $\1 Z, Z\1$. Out of
this set, $\{XX, XY, \1 Z\}$ satisfy the commutation relations, and no
other triplets which commute with these can be found, so a single qubit
can be encoded noiselessly through this channel.\qed

{\em Example 2} --- Consider the 2 qubit PIP channel with Kraus
operators proportional to $\{\1\1,YX,XY\}$. The eigenoperators with
eigenvalue $1$ are $\1\1, XY, YX$, and $ZZ$. There are no triplets
with the right commutation relations.  The eigenoperators with
eigenvalues $-1$ are $\1 Z, Z\1, XX, YY$. If we consider the $\pm1$
eigenspace, we obtain the same eigenoperators with eigenvalue 1 as the
previous example, and thus there exists a UCS consisting of a single
qubit.\qed

In the case of the previous examples, we want to map the generating
set of Pauli operators $\{\1 X,\1 Z\}$ to the generating set $\{XX,\1
Z\}$, which can be done by a controlled-NOT gate, where the second
qubit is the control, and the first qubit is the target.

The question as to whether this algorithm can find all unitarily
correctable encodings for a PIP channels remains open.


{\bf\em Verification of UCS ---} The conditions for error correction
are known to be robust against perturbations~\cite{perturbations}, and
bounds on the fidelity of the encoded states can be computed given
complete information about $\Lambda$. As argued earlier, this cannot
be done in a scalable manner. Moreover, our theorems apply
to CP maps, while it is known that there are physical maps
which are non-CP~\cite{jordan04}. For these reasons, it is important to
check what the fidelity of the encoding constructed for $\ptLambda$ is
under the action of $\Lambda$.  This can be done by twirling the
encoded channel, and estimating its average gate fidelity with the
identity channel. The estimate can be done efficiently, using a number
of experiments which is logarithmic in the number of encoded qubits,
as described in Ref.~\cite{emerson07}.

{\bf\em Discussion ---} In this paper we have shown that correctable
encodings for a quantum operation can be found by searching for
correctable encodings using the twirled version of that quantum
operation. We investigated in detail the case of channels twirled by
Pauli operators and qubit permutations, and demonstrated a simple
scheme for identifying encodings with unitary recovery
operations. Such twirled channels are important because they are
described by a polynomial number of parameters which are
experimentally accessible via a scalable protocol.  The scheme does
not require the manipulation of exponentially large matrices, 
and the performance of the constructed correctable code can be
estimated experimentally in an efficient manner. Further work is
needed to determine whether all unitarily correctable codes for PIP
channels can be found through the scheme we described. Moreover, it
would be interesting to investigate other twirls that yield more
information about the channel which may be used to find more
correctable codes.

{\bf\em Acknowledgments ---} J.E. would like to thank D. Cory for
helpful discussions. M.S. was partial supported by NSERC, MITACS
and ARO. E.M. and J.E. were partially supported by NSERC and MITACS. 
D.W.K. was partially supported by NSERC, ERA, CFI and ORF.

\end{document}